\documentclass[12pt]{iopart}

\usepackage{graphicx}
\usepackage{dcolumn}
\usepackage{caption}

\begin{document}

\title[Latest results from the
PHOBOS experiment]{Latest results from the
PHOBOS experiment}

\author{Barbara Wosiek for the PHOBOS Collaboration \footnote{For the full list
of PHOBOS authors and acknowledgments, see appendix ``Collaborations''}}

\address{Institute of Nuclear Physics Polish Academy of Sciences, Krak\'{o}w,
Poland }
\ead{barbara.wosiek@ifj.edu.pl}

%
%
B.Alver$^4$,
B.B.Back$^1$,
M.D.Baker$^2$,
M.Ballintijn$^4$,
D.S.Barton$^2$,
R.R.Betts$^6$,
A.A.Bickley$^7$,
R.Bindel$^7$,
W.Busza$^4$,
A.Carroll$^2$,
Z.Chai$^2$,
V.Chetluru$^6$,
M.P.Decowski$^4$,
E.Garc\'{\i}a$^6$,
T.Gburek$^3$,
N.George$^2$,
K.Gulbrandsen$^4$,
C.Halliwell$^6$,
J.Hamblen$^8$,
I.Harnarine$^6$,
M.Hauer$^2$,
C.Henderson$^4$,
D.J.Hofman$^6$,
R.S.Hollis$^6$,
R.Ho\l y\'{n}ski$^3$,
B.Holzman$^2$,
A.Iordanova$^6$,
E.Johnson$^8$,
J.L.Kane$^4$,
N.Khan$^8$,
P.Kulinich$^4$,
C.M.Kuo$^5$,
W.Li$^4$,
W.T.Lin$^5$,
C.Loizides$^4$,
S.Manly$^8$,
A.C.Mignerey$^7$,
R.Nouicer$^2$,
A.Olszewski$^3$,
R.Pak$^2$,
C.Reed$^4$,
E.Richardson$^7$,
C.Roland$^4$,
G.Roland$^4$,
J.Sagerer$^6$,
H.Seals$^2$,
I.Sedykh$^2$,
C.E.Smith$^6$,
M.A.Stankiewicz$^2$,
P.Steinberg$^2$,
G.S.F.Stephans$^4$,
A.Sukhanov$^2$,
A.Szostak$^2$,
M.B.Tonjes$^7$,
A.Trzupek$^3$,
C.Vale$^4$,
G.J.van~Nieuwenhuizen$^4$,
S.S.Vaurynovich$^4$,
R.Verdier$^4$,
G.I.Veres$^4$,
P.Walters$^8$,
E.Wenger$^4$,
D.Willhelm$^7$,
F.L.H.Wolfs$^8$,
B.Wosiek$^3$,
K.Wo\'{z}niak$^3$,
S.Wyngaardt$^2$,
B.Wys\l ouch$^4$\\
\vspace{3mm}
{\small
%
%
%
%
$^1$~Argonne National Laboratory, Argonne, IL 60439-4843, USA\\
$^2$~Brookhaven National Laboratory, Upton, NY 11973-5000, USA\\
$^3$~Institute of Nuclear Physics PAN, Krak\'{o}w, Poland\\
$^4$~Massachusetts Institute of Technology, Cambridge, MA 02139-4307, USA\\
$^5$~National Central University, Chung-Li, Taiwan\\
$^6$~University of Illinois at Chicago, Chicago, IL 60607-7059, USA\\
$^7$~University of Maryland, College Park, MD 20742, USA\\
$^8$~University of Rochester, Rochester, NY 14627, USA\\}

\begin{abstract}
Over the past years PHOBOS has continued to analyze the large datasets
obtained from the first five runs of the Relativistic Heavy Ion Collider (RHIC)
at Brookhaven National Laboratory. The two main analysis streams have been 
pursued. The first one aims to obtain a broad and systematic survey of global
properties of particle production in heavy ion collisions. The second class
includes the study of fluctuations and correlations in particle production.
Both type of studies have been performed for a variety of the collision 
systems,
covering a wide range in collision energy and centrality. The uniquely
large angular coverage of the PHOBOS detector and its ability to measure
charged particles down to very low transverse momentum is exploited.
The latest physics results from PHOBOS, as presented at Quark Matter 2008 
Conference, are contained in this report.

\end{abstract}

\vspace{2pc}

\section{Introduction}
PHOBOS \cite{phobos_nim} is a heavy ion experiment at the BNL RHIC collider, 
known for its 
capability of measuring charged particles over a broad angular acceptance, 
by far the largest
of all RHIC experiments. With the PHOBOS multiplicity array charged 
particles are measured in pseudorapidity range $\mid \eta\mid < 5.4$ and over
$2\pi$ in the azimuth. In addition, the unique design of the two-arm spectrometer
allows for extending the particle momentum measurements to the lowest limit
reachable at RHIC. With this detector we have collected data on 
p+p, d+Au, Cu+Cu and Au+Au collisions during the
RHIC 2000-2005 runs, covering a wide range of collision energy and 
centrality. This comprehensive data set allowed for a systematic studies of the
overall features of particle production mechanisms in nuclear and elementary
collisions, which constituted the main part of the baseline PHOBOS physics 
program. While this part is now nearly completed, our current effort is
mainly 
focused on the study of fluctuations and correlations in particle production, 
the study of which can provide deeper insight into different stages of the system
evolution. 

In this paper we briefly summarize the results on global properties of charged
particle production, including antiparticle to particle ratios and particle 
yields at very low transverse momentum ($p_T$). 
The emphasis is put on our recent 
results obtained from fluctuation and correlation studies. The dynamical 
fluctuations of the elliptic flow, corrected for the non-flow effects are
presented for Au+Au collisions at  $\sqrt{s_{NN}}$~=~200~GeV. 
The same high-statistics data set is used to investigate structures in 
the near- and away-side 
correlations with respect to high-$p_T$ trigger particle over a broad
range in $\Delta\eta$. Finally, the results from a systematic study of the 
two-particle angular
correlations in p+p, Cu+Cu and Au+Au collisions at $\sqrt{s_{NN}}$~=~200~GeV
are shown.

\section{System-size dependence of particle production}
\begin{figure}[h!]
\includegraphics[width=15cm]{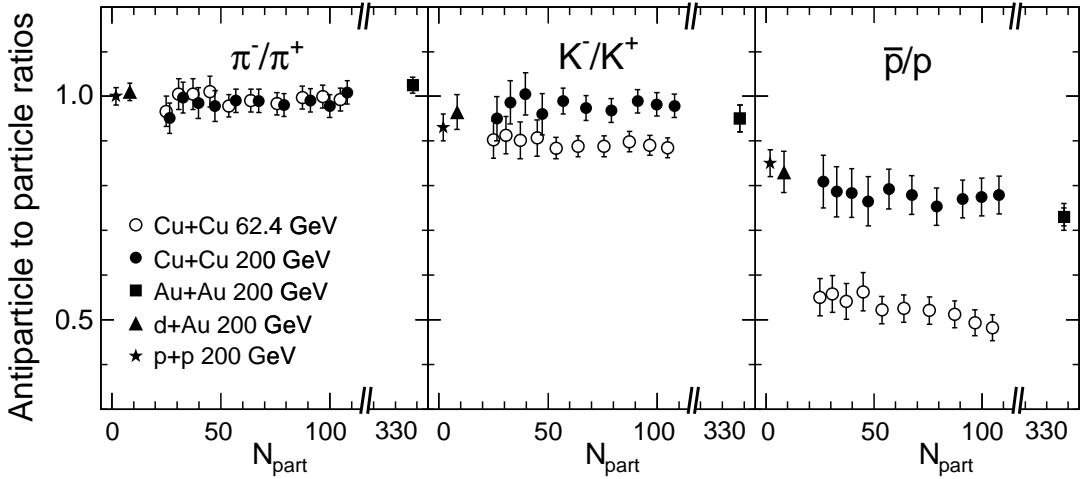}
\caption{\small Antiparticle to particle ratios for pions, kaons and protons
as a function of collision centrality for different collision systems, measured
at mid-rapidity region of $0.2 < \eta < 1.4$.}
\label{ratio}
\end{figure}

Recently  PHOBOS has completed a systematic study of the bulk 
properties of the produced particles, like total multiplicity, 
$dN_{ch}/d\eta$ and $dN_{ch}/dp_T$ distributions, particle composition and 
collective flow effects, in Cu+Cu and Au+Au collisions as a function of the
collision energy and centrality. The results, in comparison to d+Au and 
elementary p+p collisions, show that particle production in heavy ion 
collisions can be described in terms of simple scaling rules. 
Number of participant ($N_{part}$) scaling is observed for the total particle
multiplicity \cite{phobos_totmul}, net-proton yields \cite{phobos_au62} and 
also, as will be shown later, the low-$p_T$ particle yields.  
Energy and centrality dependencies of mid-rapidity yields factorize over an 
extended range of transverse momenta \cite{phobos_factpt,phobos_factmul}. 
Particle yields and directed ($v_1$) and elliptic flow ($v_2$) signals 
measured over a wide
range of high to almost central pseudorapidities show energy-independence
when viewed in the rest frame of one of the colliding nuclei 
\cite{phobos_totmul}.
The antiparticle to particle ratios show at most weak dependence
on the system size \cite{phobos_ratios}, as is illustrated by the 
compilation of PHOBOS data
presented in Figure~\ref{ratio}. 

It is also observed that $N_{part}$ unifies the measurements of the nuclear 
modification factors in Cu+Cu and Au+Au collisions \cite{phobos_raa}. 
A consistent and unified description of the elliptic flow  measured in Cu+Cu 
and Au+Au
collisions can be obtained after scaling $v_2$ with the participant 
eccentricity, $\epsilon_{part}$ \cite{phobos_flow}. 

The above observations show that the collision geometry has a major impact
on the 
dynamical evolution of the system. They also provide  a tool for extrapolating
RHIC data to the LHC energy regime, which is of particular importance since
to date no theory or model can consistently explain the scaling rules 
controlling the particle production in heavy ion collisions.

\section{Particle production at very low transverse momenta}

\begin{figure}[h!]
\begin{minipage}[t]{7.7cm}
\includegraphics[width=7.7cm]{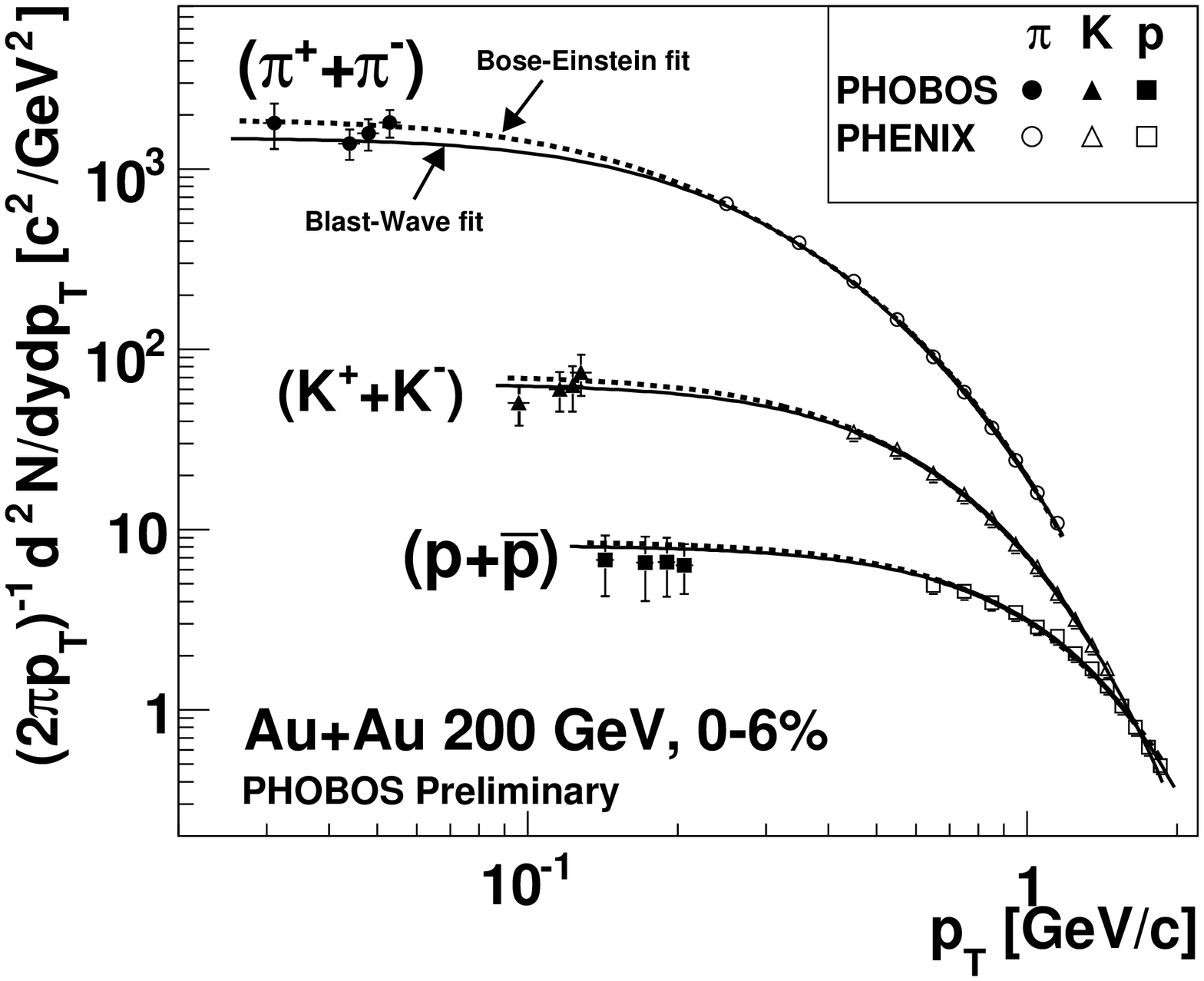}
\vspace{-1.0cm} 
\caption{\small Identified particle $p_T$ spectra near 
mid-rapidity in Au+Au collisions at 
          $\sqrt{s_{NN}}$~=~200~GeV.  }
\label{lowptyields}
\end{minipage}
\hfill
\begin{minipage}[t]{7.7cm}
\includegraphics[width=7.7cm]{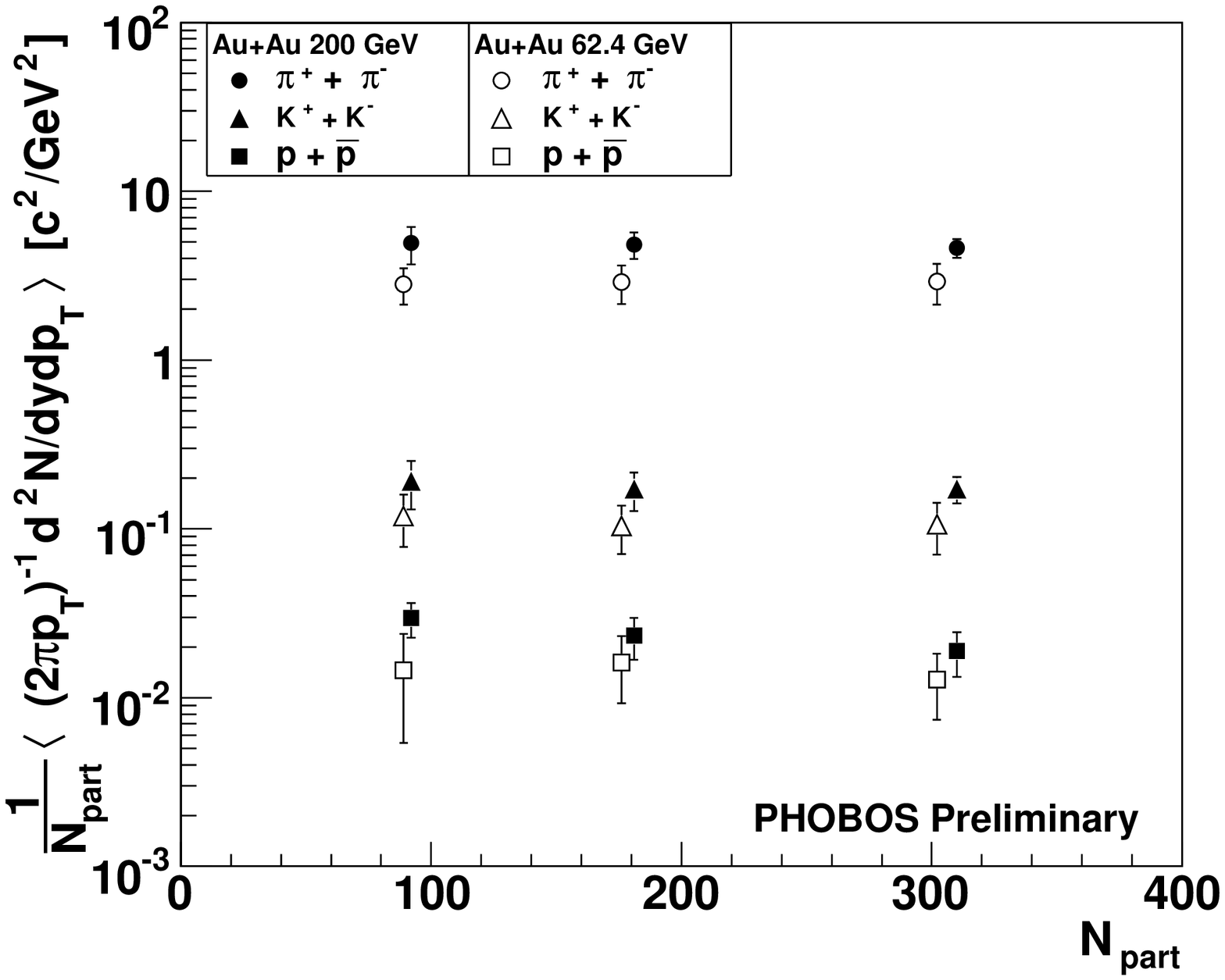} 
\vspace{-1.0cm}
\caption{\small The integrated low-$p_T$ yields for Au+Au collisions at 
200 GeV (solid~symbols) 
and 62.4 GeV (open~symbols) 
          as a function of centrality.}
\label{lowptcent}
\end{minipage}

\end{figure}

The PHOBOS spectrometer design provides  a unique capability of studying particle production
at the lowest transverse momenta accessible at RHIC. The new results 
on low-$p_T$ yields for pions, kaons and protons measured in a high-statistics 
sample
 Au+Au collisions at $\sqrt{s_{NN}}$~=~200~GeV have been shown at this 
Conference \cite{gburek}. 
Figure~\ref{lowptyields} shows the invariant low-$p_T$ yields
measured for 6\% of the most central Au+Au collisions in comparison to the 
PHENIX data at higher $p_T$ \cite{phenix_pt}. 
The low-$p_T$ yields agree with the 
extrapolations of Blast-Wave and Bose-Einstein fits to the PHENIX results.
A flattening of the shape of the $p_T$ spectra is observed, stronger for
heavier particles, which is consistent with a rapid transverse expansion of
the system. Invariant yields, integrated over the low-$p_T$ range ($0.031 - 
0.053$~GeV/c for pions, $0.105 - 0.128$~GeV/c for kaons and 
$0.143 - 0.206$~GeV/c 
for protons and antiprotons) and normalized per $N_{part}$, are shown as
a function of $N_{part}$ for Au+Au collisions at 200 and 62.4  GeV
\cite{phobos_au62} in Figure~\ref{lowptcent}. One can see that, within 
the errors, the invariant
low-$p_T$ yields scale with $N_{part}$. For more details see Ref.\cite{gburek}.

\section{Elliptic flow fluctuations}
The measured magnitude of the elliptic flow signal,  $v_2$, should reflect the 
initial spatial eccentricity of the overlap region of the colliding nuclei.
As PHOBOS has shown \cite{phobos_flow}, the participant eccentricity, 
$\epsilon_{part}$, which 
takes into account the fluctuations in the participant nucleon positions
in the overlap region, is the relevant eccentricity which drives the observed
azimuthal anisotropy. We have used the Monte Carlo Glauber (MCG) approach to
study the robustness of the participant eccentricity and also to calculate
the higher order cumulants of $\epsilon_{part}$ quantifying the event-by-event
fluctuations of the initial source eccentricity \cite{phobos_ecc}. It was
shown that the participant eccentricity
is a robust quantity, insensitive to the choice of MCG parameters and model
assumptions and that it strongly fluctuates with 
$\sigma(\epsilon_{part})/\langle\epsilon_{part}\rangle$ varying from about 
35\% 
for peripheral up to 50\% for
most central Au+Au collisions. 

\begin{figure}[h!]
\begin{minipage}[t]{7.7cm}
\includegraphics[width=7.7cm]{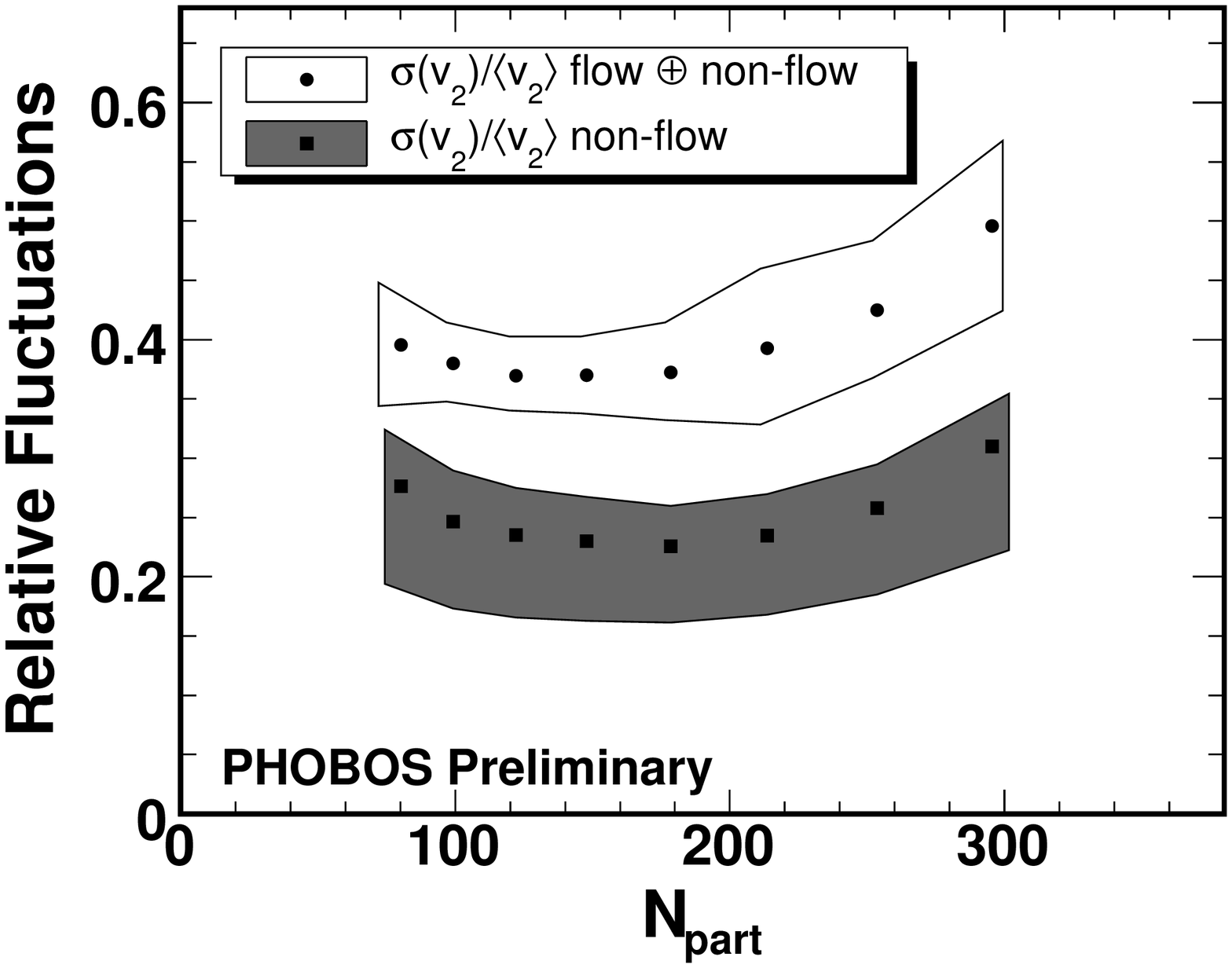}
\vspace{-1.0cm} 
\caption{\small Total measured relative flow fluctuations (circles) and the 
contribution from non-flow correlations (squares) for Au+Au collisions at 
          $\sqrt{s_{NN}}$~=~200~GeV. Bands show 90~\% C.L. systematic errors.  }
\label{flownonflow}
\end{minipage}
\hfill
\begin{minipage}[t]{7.7cm}
\includegraphics[width=7.7cm]{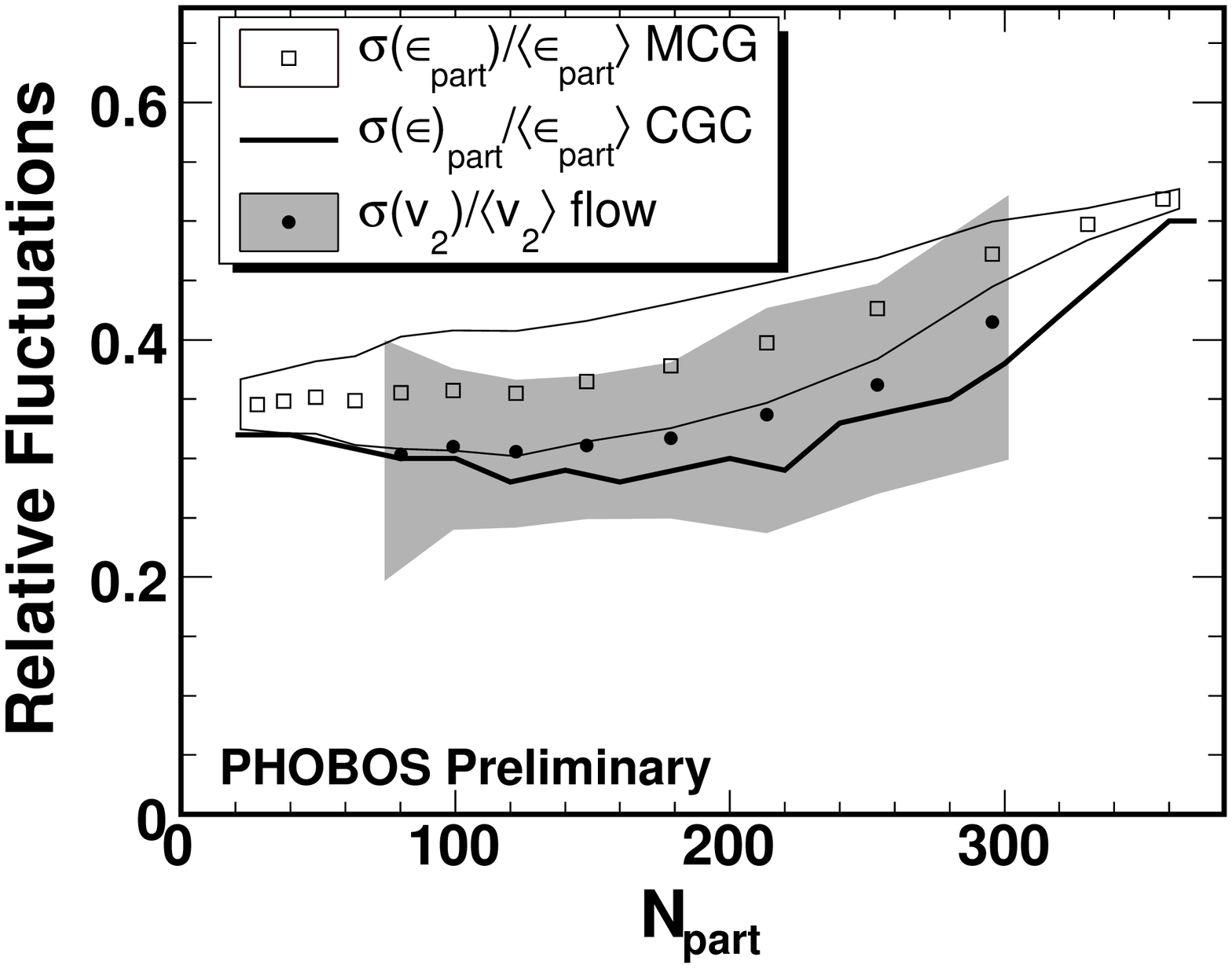} 
\vspace{-1.0cm}
\caption{\small Relative flow fluctuations, corrected for the non-flow 
effects, 
compared to the relative fluctuations of the initial eccentricity calculated 
from MC Glauber and CGC models. }
\label{fluctflow}
\end{minipage}
\end{figure}

Comparison of the $v_2$ measurements to hydrodynamic
model predictions \cite{hydro} indicate that the matter created in heavy ion 
collisions
at RHIC has properties resembling those of a perfect fluid \cite{idealfluid}.
In hydrodynamic models, the fluctuations in the shape of the initial
collision geometry should be reflected in the fluctuations of $v_2$.
PHOBOS has measured large (of the order of  $40-50$\%) 
relative fluctuations of the elliptic
flow, $\sigma(v_2)/\langle v_2 \rangle$ for Au+Au collisions at  
$\sqrt{s_{NN}}$~=~200~GeV 
\cite{flowfluct}.
The main  difficulty in this measurement 
is to disentangle dynamical fluctuations of $v_2$ and non-flow correlations due,
for example, to resonance decays or jet production. PHOBOS has developed a
new method to extract the non-flow component from the data. The method relies 
on the determination of the second Fourier coefficient of two-particle angular
correlations  measured over a broad range of $\Delta\eta$, thanks to the large
acceptance of the PHOBOS detector. This coefficient, 
$v_{2}^{2}(\eta_1,\eta_2)$, contains contributions from genuine flow 
correlations, $v_2(\eta_1) \times v_2(\eta_2)$,
and non-flow correlations, $\delta(\eta_1,\eta_2)$. At $\Delta\eta > 2$, 
we assume that the
non-flow component is small. In fact a small non-flow effect at 
$\Delta\eta > 2$ was estimated from HIJING simulations and accounted for. 
Thus, performing the
fit to the second Fourier coefficient of two-particle angular
correlations allows us to determine the flow component, 
$v_2(\eta_1) \times v_2(\eta_2)$, and then estimate the non-flow contribution
by subtracting flow component from  
$v_{2}^{2}(\eta_1,\eta_2)$. The contribution from non-flow correlations
to $\sigma(v_2)$ is $\sqrt{\langle \delta \rangle/2}$ \cite{sigmanonflow}.
 Figure~\ref{flownonflow} shows the measured total 
relative $v_2$ fluctuations and the non-flow contribution to these 
fluctuations. The non-flow contribution extracted from the data is large, of
the order of $25 - 30$\%. The relative flow fluctuations, after subtracting the
non-flow contribution, are shown in 
Figure~\ref{fluctflow} and compared to the relative 
fluctuations of the initial eccentricity calculated from MCG
 \cite{phobos_ecc}
and Color Glass Condensate (CGC) \cite{cgc} models. It can be seen that 
the magnitude of
the relative flow fluctuations agree, within the errors,  with both MCG
and CGC calculations of the fluctuations in initial source eccentricity, 
leaving essentially no room for other, later-stage contributions. For more 
details see Ref.~\cite{Alver}.

\section{High-$p_T$ triggered two-particle correlations}

\begin{figure}[h!]
\begin{minipage}[t]{7.7cm}
\includegraphics[width=7.7cm]{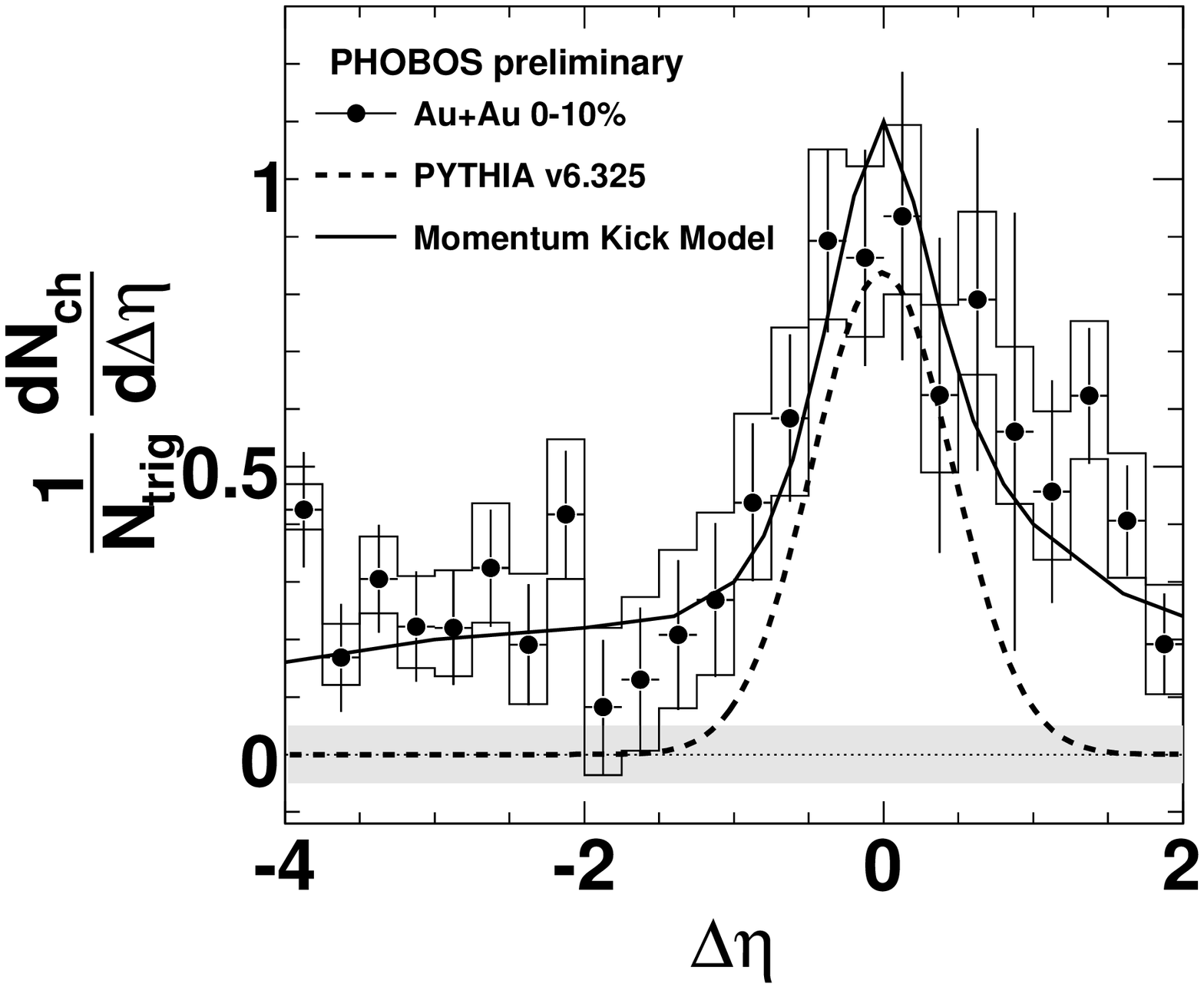}
\vspace{-1.0cm} 
\caption{\small The correlated near-side yield for 0-10\% central Au+Au 
collisions compared to p+p PYTHIA simulations (dashed line)
 and momentum kick model
predictions (solid line).  }
\label{wenger1}
\end{minipage}
\hfill
\begin{minipage}[t]{7.7cm}
\includegraphics[width=7.7cm]{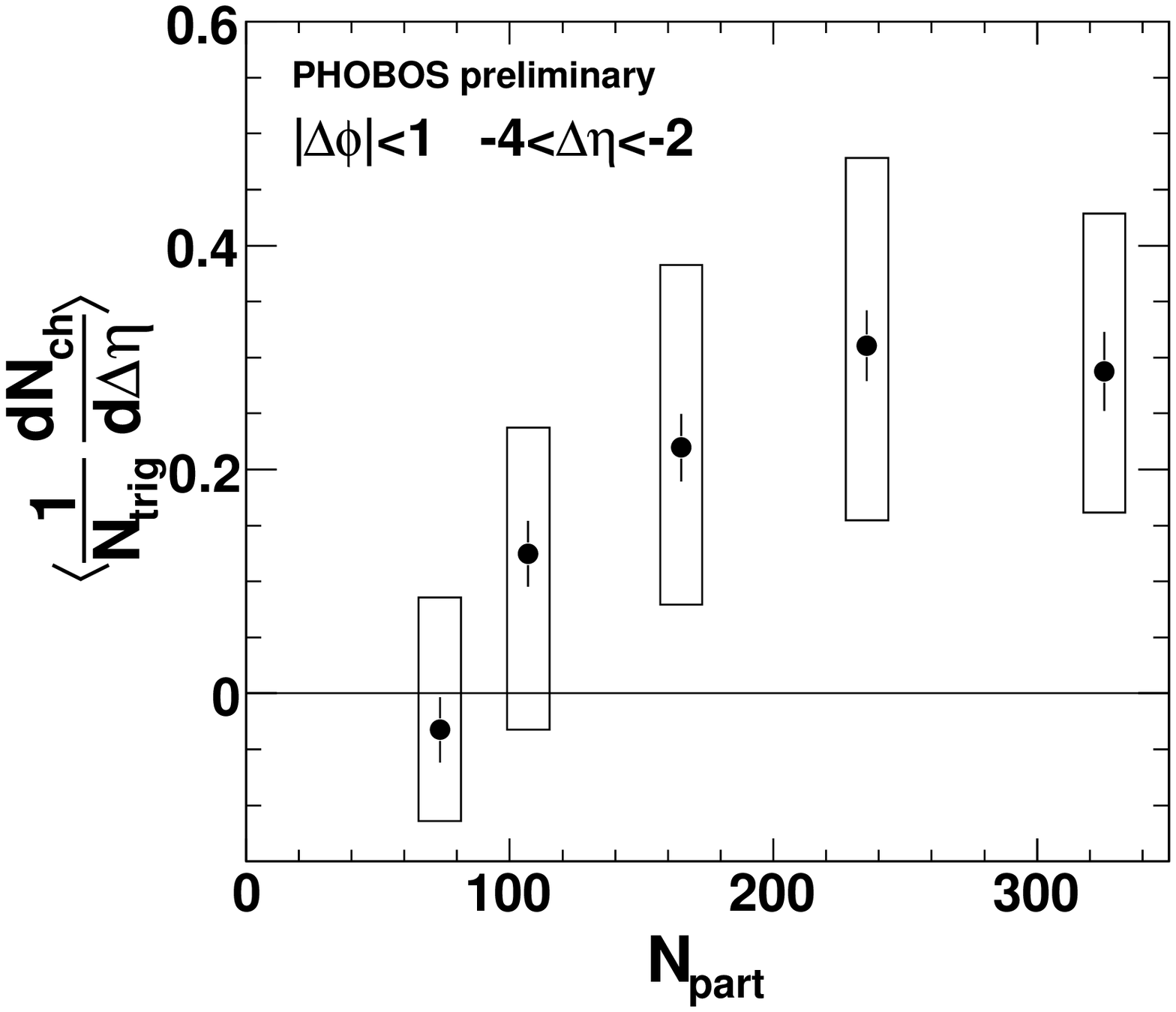} 
\vspace{-1.0cm}
\caption{\small The near-side yield averaged over $ -4 < \Delta\eta < -2$ as
a function of $N_{part}$.  }
\label{wenger2}
\end{minipage}
\end{figure}

Measurements of correlations with respect to the high-$p_T$ trigger
particle allow us to study the medium response to energetic partons produced
in early hard scattering processes and then propagated through the dense
medium created in heavy ion collisions. Here we report the first measurements
of the $\Delta\eta$ and $\Delta\phi$ correlations between the triggered charged
particle with $p_T > 2.5$~GeV/c and all associated hadrons recorded in
the PHOBOS Octagon detector \cite{phobos_nim}. The analysis utilizes the 
extensive coverage in 
pseudorapidity ($\mid \eta \mid < 3$) and azimuthal angle ($\Delta \phi =2\pi$)
of the Octagon allowing for the study of both short- and long-range 
correlations
in $\Delta\eta$. A particular emphasis is put on testing the presence of 
long-range
correlations at near- and away-sides over the uniquely broad acceptance in 
$\Delta\eta$ 
($-4 < \Delta\eta < 2$). The correlated yields per trigger particle, with
the elliptic flow effects subtracted, are 
calculated for Au+Au collisions at  $\sqrt{s_{NN}}$~=~200~GeV and compared to 
the correlated particle production in p+p events simulated with PYTHIA. The two dimensional correlated yield, 
$\frac{1}{N_{trig}}\frac{d^2N_{ch}}{d\Delta\eta d\Delta\phi}$,
in central Au+Au collisions exhibits a much broader away-side peak  in 
$\Delta\phi$ as compared to p+p correlations. The near-side yield in central
Au+Au collisions
shows a
clear jet-like peak near $\Delta\phi \approx 0$ as well as pronounced correlations extending over
a broad $\Delta\eta$  range (the so called ridge structure). 
In Figure~\ref{wenger1} the correlated near-side yield (integrated over 
$\mid\Delta\phi\mid <1$) is plotted as a function of $\Delta\eta$ for 10\% 
of the
most central Au+Au collisions. A prominent ridge effect is clearly visible
up to  $\Delta\eta = -4$ in contrast to p+p PYTHIA events, where the structure
is absent. The momentum kick model \cite{wong} is able to
quantitatively describe this structure. Centrality dependence of the
 near-side yield, averaged over 
$ -4 < \Delta\eta < -2$, is shown in Figure~\ref{wenger2}. One can see
that the ridge effect weakens with decreasing collision centrality and 
almost completely disappears for $N_{part}$ less than  about 80. For more details
of this analysis see Ref.~\cite{wenger}. 

\section{Two-particle angular correlations}

\begin{figure}[h!]
\begin{minipage}[t]{7.7cm}
\includegraphics[width=7.7cm]{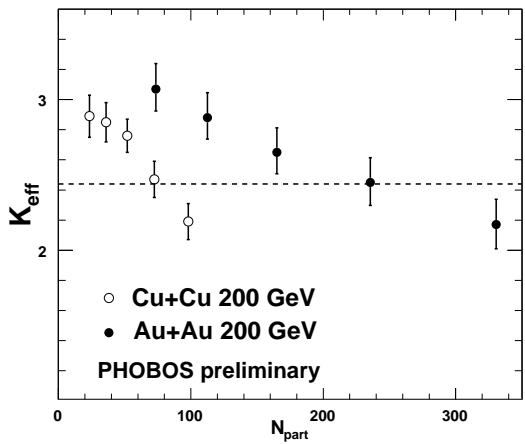}
\vspace{-1.0cm} 
\caption{\small Effective cluster size as a function of $N_{part}$ for
Cu+Cu and Au+Au collisions at  $\sqrt{s_{NN}}$~=~200~GeV. 
Dashed line denotes the value of 
$K_{eff}$
measured in p+p 
collisions.  }
\label{wei_npart}
\end{minipage}
\hfill
\begin{minipage}[t]{7.7cm}
\includegraphics[width=7.7cm]{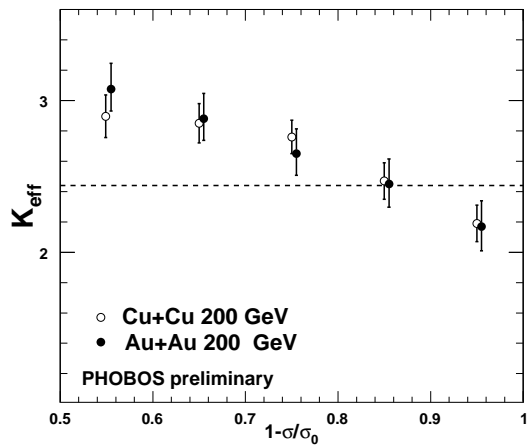} 
\vspace{-1.0cm}
\caption{\small The same as in Figure~\ref{wei_npart} but plotted against the
fractional cross-section.  }
\label{wei_cross}
\end{minipage}
\end{figure}

The study of the correlations between the final-state particles should provide
additional information on the underlying mechanism of particle production at
freeze-out. PHOBOS has performed systematic studies of two-particle angular
correlations in ($\Delta\eta, \Delta\phi$) in elementary p+p collisions 
\cite{phobos_ppcorr}
as well as Cu+Cu collisions  \cite{phobos_cucucorr}.  Recently these studies
have been also carried out for Au+Au collisions at  $\sqrt{s_{NN}}$~=~200~GeV.
The broad coverage of the PHOBOS Octagon detector is fully utilized in these
studies. The multiplicity-independent two-particle correlation function,
$R(\Delta\eta,\Delta\phi)$, is calculated using the method detailed in 
\cite{phobos_ppcorr}. We have concentrated on studying the short-range
correlations by projecting the two-dimensional correlation function
into a one-dimensional correlation function $R(\Delta\eta)$. $R(\Delta\eta)$
exhibits clear short-range correlations which can be described using a 
simple cluster model. The one-dimensional correlation function is fitted to a 
parameterization derived in  an independent cluster model \cite{cluster_model}
in order to extract the effective cluster size, $K_{eff}$. 
Figure~\ref{wei_npart} shows the effective cluster size as a function of
$N_{part}$ for  $\sqrt{s_{NN}}$~=~200 GeV Cu+Cu and Au+Au collisions. 
For both collision systems
the cluster size decreases with increasing centrality. Furthermore, for the
same $N_{part}$, clusters in Au+Au collisions are larger than in the Cu+Cu system.
Interestingly, this dependence on the size of the colliding nuclei 
disappears after plotting
the same data as a function of fractional cross-section as shown in
Figure~\ref{wei_cross}. More differential studies of the effective size of  
near- and away-side clusters have also been performed. It is observed that
away-side clusters are smaller and decrease more rapidly with increasing
centrality than near-side clusters. This observation can be qualitatively
explained by assuming that near-side clusters are preferentially
produced close to the surface of the emission zone, while away-side
clusters seem to propagate through the medium. 
 For more details on the two-particle correlation study see
\cite{wei}.

\section{Summary and outlook}
PHOBOS is continuing to provide interesting and unique data on particle
production in different collision systems in the RHIC energy range. 
The uniqueness of the PHOBOS
data is owed to the distinct features of the PHOBOS detector: its extensive
acceptance for charged particle measurements and capability of measuring
particles  at very low $p_T$. 
Comprehensive results on global features of the particle production are only
briefly summarized in this report. 
The measurements of low-$p_T$ yields, for the first
time performed as a function of centrality, are presented for Au+Au collisions
at $\sqrt{s_{NN}}$~=~200~GeV. No anomalous enhancement of pion production is
observed at very low $p_T$. The spectral shapes for heavier  particles show
flattening effects, consistent with a strong transverse expansion of the 
system. 

In this report we have presented a wealth
of results on the fluctuations and correlation measurements from PHOBOS.
New results on event-by-event elliptic flow fluctuations, corrected for the
non-flow effects extracted from data, are shown for Au+Au collisions at
$\sqrt{s_{NN}}$~=~200~GeV. Although non-flow correlations contribute
significantly to the measured flow fluctuations, the corrected relative flow 
fluctuations are large, with a magnitude in agreement with calculations of
fluctuations in the participant eccentricity. These results indicate that system
thermalizes very rapidly and the initial-state event-by-event source 
fluctuations are efficiently converted into final-state momentum fluctuations.

The PHOBOS measurements of correlations between the trigger high-$p_T$ 
particle and 
associated particles with no $p_T$ cut imposed, allow for studying the
ridge structure in the near-side correlations over the longitudinal extent 
larger than accessible in other RHIC experiments. The results for
Au+Au collisions at $\sqrt{s_{NN}}$~=~200~GeV show that the ridge structure
of the near-side correlations persists up to the limit of our acceptance, i.e.
$\Delta\eta = 4$. With decreasing collision centrality the structure becomes
less pronounced, eventually disappearing at $N_{part}$ of about 80. These 
results provide valuable constraints on models aimed to describe jet 
propagation through a dense medium.

The systematic studies of two-particle angular correlations for p+p, Cu+Cu and
Au+Au collisions at  $\sqrt{s_{NN}}$~=~200~GeV show that at freeze-out particles tend to be 
produced in clusters with a non-trivial centrality dependence observed in
nucleus-nucleus collisions. Furthermore, an unexpected system size dependence
is seen. These observations provide a challenge  to models describing
particle production in heavy ion collisions. 

The PHOBOS will continue the analysis of the existing comprehensive dataset.
The studies of the low-$p_T$ particle production as well as of elliptic flow
fluctuations and triggered correlations will be extended to other systems
and energies. Although PHOBOS has accomplished its initial goal, the future 
studies should still provide results improving our understanding of the
physics of heavy ion collisions. 

\vspace*{3mm}
%
%
%
%
{\small This work was partially supported by U.S. DOE grants 
DE-AC02-98CH10886,
DE-FG02-93ER40802, 
DE-FG02-94ER40818,  
DE-FG02-94ER40865, 
DE-FG02-99ER41099, and
DE-AC02-06CH11357, by U.S. 
NSF grants 9603486, 
0072204,            
and 0245011,        
by Polish MNiSW grant N N202 282234 (2008-2010),
by NSC of Taiwan Contract NSC 89-2112-M-008-024, and
by Hungarian OTKA grant (F 049823).}

\end{document}